\documentclass[12pt]{article}
\usepackage{bm}
\newcommand{\nn}{\nonumber\\}

\newcommand{\abs}[1]{\left| #1 \right|}

\newcommand{\Q}{Q_{\rm B}}
\newcommand{\hQ}{\hat{Q}_{\rm B}}
\newcommand{\heta}{\hat{\eta}}
\newcommand{\hLambda}{\hat{\Lambda}}
\newcommand{\id}{\bm{1}}

\newcommand{\Bra}{\Big< \hspace{-1mm}\Big<}
\newcommand{\Ket}{\Big> \hspace{-1mm}\Big>}
\renewcommand{\thepage}{}
\makeatletter
\@addtoreset{equation}{section}
\renewcommand{\theequation}{\thesection.\@arabic\c@equation}
\makeatother
\renewcommand{\thefootnote}{\fnsymbol{footnote}}
\begin{document}
\begin{titlepage}
\title{
\vspace*{-4ex}
\hfill
\begin{minipage}{3.5cm}
\normalsize KEK-TH-1045\\
\normalsize hep-th/0510224
\end{minipage}\\
\vspace{4ex}
\bf Analytical Tachyonic Lump Solutions in Open Superstring Field Theory 
\vspace{5ex}}
\author{Isao {\sc Kishimoto}$^{1,}$\footnote{E-mail address:
ikishimo@post.kek.jp}\ \ and\ \  
Tomohiko {\sc Takahashi}$^{2,}$\footnote{E-mail address:
tomo@asuka.phys.nara-wu.ac.jp}\\
\vspace{2ex}\\
$^1${\it High Energy Accelerator Research Organization (KEK),}\\
{\it Tsukuba, Ibaraki 305-0801, Japan}\\
$^2${\it Department of Physics, Nara Women's University,}\\
{\it Nara 630-8506, Japan}}
\date{October, 2005}
\maketitle
\vspace{7ex}

\begin{abstract}
\normalsize
\baselineskip=19pt plus 0.2pt minus 0.1pt

We construct a classical solution in the GSO$(-)$ sector
in the framework of a Wess-Zumino-Witten-like open superstring 
field theory on a non-BPS D-brane.
We use an $su(2)$ supercurrent, which is obtained by
compactifying a direction to a circle with the critical radius,
in order to get analytical tachyonic lump solutions to the equation of
 motion. By investigating the action expanded around a solution
we find that it represents a deformation from a non-BPS D-brane
to a D-brane-anti-D-brane system
at the critical value of a parameter which is contained in
classical solutions.
Although such a process was discussed in terms of boundary 
conformal field theory before, our study is based on open superstring
 field theory including interaction terms.

\end{abstract}
\end{titlepage}

\renewcommand{\thepage}{\arabic{page}}
\renewcommand{\thefootnote}{\arabic{footnote}}
\setcounter{page}{1}
\setcounter{footnote}{0}
\baselineskip=19pt plus 0.2pt minus 0.1pt
%

\section{Introduction}

Analytic classical solutions have been found in open superstring
field theory \cite{KT2,Lechtenfeld:2002cu,Kluson:2002kk} on BPS D-branes 
formulated in terms of the Wess-Zemuno-Witten 
(WZW) like action \cite{rf:SSFT1,rf:SSFT2}.\footnote{A string field
theory around the solutions in ref.~\cite{Lechtenfeld:2002cu} was
analyzed in refs.~\cite{Kling:2002ht,Kling:2002vi}} The solutions in
ref.~\cite{KT2} are constructed from supercurrents, 
ghost  fields and the identity string field. The characteristic features
of the solution are 
its correspondence to the marginal deformation generated by the
supercurrent and a well-defined Fock space expression of the
solution. 

In open bosonic string field theory, similar classical solutions have
been constructed by using
currents \cite{KT2,rf:TT1,rf:TT2,rf:KTZ,Kluson_M,Kluson_02}. 
They also correspond to marginal
deformations and have 
a well-defined Fock space expression. Unfortunately, the vacuum energy
of the bosonic solution is provided as a kind of indefinite quantities.
In the absence of appropriate regularization,
we have nothing else to do but evaluate it by indirect calculation.
However, the remarkable feature of the supersymmetric solutions is that
their vacuum energy vanishes exactly by a direct calculation \cite{KT2} as
expected from their correspondence to marginal deformations \cite{rf:LMD}. 
The supersymmetric 
case may provide a clue for solving the vacuum energy problem in the
bosonic case.

Even on non-BPS D-branes, we can formulate string field
theory in terms of the WZW like action \cite{rf:SSFT2,BSZ}. Since non-BPS
D-branes have 
tachyonic modes in the GSO$(-)$ sector, the theory enables us to
investigate D-brane decay processes. 
A tachyonic lump solution, for instance, describes a deformation
from a non-BPS D-brane to a  D-brane-anti-D-brane pair. 
Actually, several analyses were performed by using the level
truncation scheme \cite{BSZ,rf:DSR,rf:IN,rf:Ohmori}.  
If one of the directions is compactified on
the circle with the critical radius, the above process is realized by a
marginal deformation \cite{Sen:1998ex}. Accordingly, we have only to
extend the solution 
on the BPS D-brane to the non-BPS case in order to construct the solution
corresponding to the tachyonic lump. 
In this paper, we will construct an analytical solution of this
tachyonic lump.

In open superstring field theory on a single non-BPS D-brane,
the action of the NS sector string field
is given by \cite{rf:SSFT2,BSZ}
\begin{eqnarray}
\label{Eq:action}
 S[\hat{\Phi};\hQ]=\frac{1}{4g^2} \Bra
(e^{-\hat{\Phi}} \hQ e^{\hat{\Phi}})(e^{-\hat{\Phi}}\heta_0
e^{\hat{\Phi}})
-\int_0^1\!dt\,(e^{-t\hat{\Phi}}\partial_t e^{t\hat{\Phi}})
\left\{(e^{-t\hat{\Phi}}\hQ e^{t\hat{\Phi}}),\,
(e^{-t\hat{\Phi}}\heta_0 e^{t\hat{\Phi}})\right\}
\Ket,
\end{eqnarray}
where $\hat{\Phi}$ denotes a string field of NS sector which corresponds
to a vertex operator of ghost number 0 and picture number 0 in the
conformal field theory (CFT). In order to incorporate GSO$(-)$ sector
into the theory on a BPS D-brane, we have to introduce internal
Chan-Paton factors: 
\begin{eqnarray}
 \hat{\Phi}=\Phi_+\otimes \id 
           +\Phi_-\otimes \sigma_1,
\end{eqnarray}
where the subscript $+$ ($-$) implies that the corresponding vertex
operator 
is in the GSO$(+)$ (GSO$(-)$) sector. The operators $\hQ$
and $\heta_0$ are defined as
\begin{eqnarray}
\label{Eq:BRStensor}
 \hQ = \Q \otimes \sigma_3,\ \ \ 
  \heta_0 = \eta_0 \otimes \sigma_3,
\end{eqnarray}
where $\Q$ and $\eta_0$ are the ordinary operators without cocycle
factors. 
The bracket $\langle\!\langle\cdots \rangle\!\rangle$ is
defined by a CFT correlator in the large Hilbert space and
a trace over internal Chan-Paton matrices. 
The action is invariant under the infinitesimal gauge
transformation,
$ \delta e^{\hat{\Phi}}=(\hQ\delta\hLambda)*e^{\hat{\Phi}}+
e^{\hat{\Phi}}*(\heta_0\delta\hLambda')$,
where $\delta\hLambda$ and $\delta\hLambda'$ are infinitesimal
parameters.
Variating the action (\ref{Eq:action}), we can derive the
equation of motion to be 
\begin{eqnarray}
\label{Eq:eom}
\heta_0(e^{-\hat{\Phi}}*\hQ e^{\hat{\Phi}})=0.
\end{eqnarray}
This is the equation to be solved in this paper.
For details of the 
definition, see for instance ref.~\cite{BSZ}.

This paper is organized as follows. In section 2, we will construct a
tachyonic lump solution. In the bosonic case at the critical radius,
there is an $su(2)$ current algebra which is useful to discuss descent
relations of bosonic D-branes and to construct analytical lump
solutions. At first, we find that a similar $su(2)$ supercurrent algebra
exists even in the theory on a non-BPS D-brane. Using this supercurrent,
we can solve the equation of motion and find an analytic lump solution.
The vacuum energy of the resulting solution vanishes exactly as well as
the BPS case.
In section 3, we will discuss the theory expanded around the tachyonic
lump solution. 
To interpret physical meanings of the expanded theory, fermionization of
the compactified direction plays a key role, that is used to discuss a
tachyonic lump in the context of boundary conformal field theory
\cite{MajSen, Sen:1998ex}.
Finally, we find that at the critical value of the solution the expanded
theory is equivalent to the theory on a D-brane-anti-D-brane
pair. Although this result is expected from boundary conformal field
theory, we will provide a complete proof including interaction terms
based on the analytic classical solution to eq.~(\ref{Eq:eom}) in 
open superstring field theory. In section 4 we conclude with a brief
summary and open problems.

\section{Tachyonic lump solutions}

\subsection{$su(2)$ supercurrent algebra}

We compactify one of tangential directions to the brane on a circle of 
radius $R=\sqrt{2\alpha'}$. We take it as the 9-th direction and write
the string coordinate as $X^9(z,\bar{z})=(X^9(z)+X^9(\bar{z}))/2$ and
its supersymmetric partner as $\psi^9(z)$. The operator product
expansions (OPEs) of these fields are given by $X^9(y)X^9(z)\sim
-2\alpha'\log(y-z)$ and $\psi^9(y)\psi^9(z)\sim 1/(y-z)$.

Similarly to the bosonic case, we can find a level
one $su(2)$ supercurrent algebra at the critical radius, although the
critical radius in the superstring case is inequivalent to that of
bosonic case. The three supercurrents are given by
\begin{eqnarray}
\label{eq:J1v2}
 {\bf J}^1(z,\theta)&=&\sqrt{2}\sin\left(
X^9\over \sqrt{2\alpha'}\right)\!(z)c_1+\theta
\sqrt{2}\psi^9\cos\left(
X^9\over \sqrt{2\alpha'}\right)\!(z)c_2\,,
\\
\label{eq:J2v2}
 {\bf J}^2(z,\theta)&=&\sqrt{2}\cos\left(
X^9\over \sqrt{2\alpha'}\right)\!(z)c_1+\theta
(-\sqrt{2})\psi^9\sin\left(X^9\over \sqrt{2\alpha'}\right)\!(z)c_2\,,\\
\label{eq:J3v2}
 {\bf J}^3(z,\theta)&=&\psi^9(z)c_3+\theta{i\over
  \sqrt{2\alpha'}}\partial X^9(z)\,.
\end{eqnarray}
Here, we have introduced the cocycle factors $c_i$ defined as
\begin{eqnarray}
 c_3^2=1,~~~~c_ic_j=\delta_{ij}+i\epsilon_{ijk}c_k~~~~i,j,k=1,2,3,
~~~\epsilon_{123}=+1\,,
\end{eqnarray}
where $\epsilon_{ijk}$ is the totally antisymmetric tensor.
Writing ${\bf J}^a(z,\theta)\equiv\psi^a(z)+\theta J^a(z)~~(a=1,2,3)$,
we obtain the following current algebra,
\begin{eqnarray}
&&\psi^a(y)\psi^b(z)\sim (y-z)^{-1}\delta_{ab}\,,\\
&&J^a(y)\psi^b(z)\sim (y-z)^{-1}(-i\epsilon_{abc}\psi^c(z))\,,\\
&&J^a(y)J^b(z)\sim
 (y-z)^{-2}\delta_{ab}+(y-z)^{-1}(-i\epsilon_{abc}J^c(z))\,.
\end{eqnarray}
This is the same as $su(2)$ supercurrent algebra obtained by substituting 
$\Omega^{ab}=2\delta_{ab}$ and $f^{ab}_{~~c}=-i\epsilon_{abc}$
in eqs.~(3.1--3.3) in ref.~\cite{KT2}.

From these supercurrents, we can construct the energy-momentum tensor
by using the Sugawara method. First, we can find the following equations,
\begin{eqnarray}
&&-:\psi^1\partial\psi^1:(z)=-{1\over 4\alpha'}(\partial X^9)^2(z)-\cos\left(
2X^9\over \sqrt{2\alpha'}
\right)(z)\,,\\
&&-:\psi^2\partial\psi^2:(z)=-{1\over 4\alpha'}(\partial X^9)^2(z)+\cos\left(
2X^9\over \sqrt{2\alpha'}
\right)(z)\,,\\
&&:J^1J^1:(z)=-\psi^9\partial\psi^9(z)+\cos\left(
2X^9\over \sqrt{2\alpha'}
\right)(z)-{1\over 4\alpha'}(\partial X^9)^2(z)\,,\\
&&:J^2J^2:(z)=-\psi^9\partial\psi^9(z)-\cos\left(
2X^9\over \sqrt{2\alpha'}
\right)(z)-{1\over 4\alpha'}(\partial X^9)^2(z)\,,\\
&&:\psi^a\psi^b:(z)=-i\epsilon_{abc}J^c(z)~~~~~(a\ne b)\,,\\
&&:\psi^aJ^b:(z)\,-:J^a\psi^b:(z)=i\epsilon_{abc}\partial\psi^c(z)\,,
~~~~(a\ne b)\,,\\
&&:J^1\psi^1:(z)=\,:J^2\psi^2:(z)=\,:J^3\psi^3:(z)={i\over
 \sqrt{2\alpha'}}\psi^9\partial X^9(z)c_3\,.
\end{eqnarray}
Then, we obtain the energy-momentum tensor and the world-sheet
supercurrent as\footnote{We have used the definition in ref.~\cite{KT2}.}
\begin{eqnarray}
 T^9(z)&=&{1\over 2}:(J^aJ^a+\partial\psi^a\psi^a):(z)
-{i\over
6}\epsilon_{abc}:(J^a:\psi^b\psi^c:+\psi^a:(\psi^bJ^c-J^b\psi^c):):(z
)\nonumber\\
\label{Eq:emtensor}
&=&-{1\over 4\alpha'}(\partial X^9)^2(z)-{1\over
2}\psi^9\partial\psi^9(z)\,,\\
G^9(z)&=&:J^a\psi^a:(z)-{i\over
 3}\epsilon_{abc}:\psi^a:\psi^b\psi^c::(z)
\label{Eq:scurrent}
={i\over
 \sqrt{2\alpha'}}\psi^9\partial X^9(z)c_3\,.
\end{eqnarray}
Here, we should note that the world-sheet supercurrent $G^9(z)$ contains
the cocycle factor $c_3$.
In the existence of cocycle factors, the operators, $T^9(z)$, $G^9(z)$,
$\psi^a(z)$ and $J^a(z)$, satisfy a superconformal current algebra with
$c=3/2$. 

To incorporate the GSO$(-)$ states into a string field, we have to
introduce internal Chan-Paton indices as in ref.~\cite{rf:SSFT2,BSZ}. In
the theory on a non-BPS D-brane, fermionic operators like the BRS
charge 
are tensored with a Pauli matrix as seen in eq.~(\ref{Eq:BRStensor}).
Since the world-sheet supercurrent $G^9(z)$ is a fermionic operator and
it is tensored with $c_3$, we identify the cocycle factors $c_i$
with Pauli matrices representing the internal Chan-Paton factors:
\begin{eqnarray}
\label{eq:c123}
 c_3=\sigma_3,\ \ \ c_1=\sigma_2,\ \ \ c_2=-\sigma_1.
\end{eqnarray}

We note that the
$SU(2)$ symmetry does not realized on a non-BPS D-brane
in spite of the fact that the $su(2)$ supercurrent algebra exists in the
theory. In fact,
$J_0^a=\oint \frac{dz}{2\pi i}J^a(z)$ is not to be a derivation 
with respect to the star product 
although $[\hat{Q}_{\rm B},\,J_0^a]=[\heta_0,\,J_0^a]=0$.
Because we find that
$J_0^a(\hat{\Psi}_1*\hat{\Psi}_2)=(J_0^a\hat{\Psi}_1)*\hat{\Psi}_2 
+((-1)^{\hat{F}+\hat{n}}\hat{\Psi}_1)*(J_0^a\hat{\Psi}_2)$ 
for $a=1,2$ due to cocycle factors
and quantized momentum along the 9-th
direction, where $\hat{F}$ and $\hat{n}$ are operators counting fermion
number and the 9-th momentum as defined later by (\ref{eq:F}) and
(\ref{eq:n1}). But, at the same time, 
$J_0^3(\hat{\Psi}_1*\hat{\Psi}_2)=(J_0^3\hat{\Psi}_1)*\hat{\Psi}_2
+\hat{\Psi}_1*(J_0^3\hat{\Psi}_2)$. Thus, the $SU(2)$ symmetry is broken 
to $U(1)$ by the interaction terms in the action (\ref{Eq:action}).

\subsection{Classical solutions via supercurrents}

In ref.~\cite{KT2}, we have constructed analytic classical solutions
in the theory on BPS D-branes by means of supercurrent algebra.
Now that we possess the supercurrent algebra including GSO$(-)$ sector,
we can apply the same method to the theory on the non-BPS
D-branes. Taken $\mathbf{J}^1(z,\theta)$ as the supercurrent, the
classical solution is given by 
\begin{eqnarray}
&&\hat{\Phi}_0=-\tilde{V}_L(F)I\,,
\label{eq:solution_gen1}\\
&&\tilde{V}_L(F)=\int_{C_{\rm left}}{dz\over 2\pi i}
F(z){\tilde v}(z),\\
\label{Eq:vtilde}
&&\tilde{v}(z)=\frac{1}{\sqrt{2}}\,c\xi e^{-\phi}(z)\otimes \sigma_3
 \times  \psi^1(z), 
\end{eqnarray}
where $I$ is the identity string field and
$C_{\rm left}$ denotes a counter-clockwise path along a half
of the unit circle, i.e., $-\pi/2<\sigma<\pi/2$ for $z=e^{i\sigma}$.
$F(z)$ is a function on the unit circle $\abs{z}=1$ satisfying
$F(-1/z)=z^2 F(z)$ \cite{KT2}.\footnote{Under this condition,
$F(z)$ cannot be a non-zero constant.}
We must impose an additional constraint on $F(z)$ due to 
the reality condition of the string field as in ref.~\cite{KT2}. 
The cocycle factor $\sigma_3$
should be attached  in $\tilde{v}(z)$ since the ghost factor $c\xi
e^{-\phi}(z)$ is Grassmann odd.\footnote{We note that $e^{q\phi}$
($q$\,:\,odd) is a fermionic operator. 
More precisely, we need a cocycle factor to represent statistical
property of the operator.} Substituting $\psi^1(z)$ of
eq.~(\ref{eq:J1v2}) into eq.~(\ref{Eq:vtilde}), the operator
$\tilde{v}(z)$ is rewritten as\footnote{We have adjusted $c_1=\sigma_2,\
c_2=-\sigma_1$ in eq.~(\ref{eq:c123}) 
so that $\tilde{v}(z)$ has the cocycle factor $\sigma_1$.
If we choose $\psi^2$ instead of $\psi^1$ in eq.~(\ref{Eq:vtilde}),
we get cosine-type solution. These sine and cosine type solutions are
related by $U(1)$-symmetry, which is generated by $J^3_0$.}
\begin{eqnarray}
\tilde{v}(z)=-i\,c\xi e^{-\phi}\sin\left(
\frac{X^9(z)}{\sqrt{2\alpha'}}\right)\otimes \sigma_1.
\end{eqnarray}
It turns out that this classical solution represents a non-trivial
configuration of the GSO$(-)$ string field. Since the GSO$(-)$ states
include a tachyonic mode, this solution can be regarded as a kind of
tachyonic lump solutions.

Now, we can easily find that the equation
of motion actually holds. First, we define the operator $V_L(g)$ as 
\begin{eqnarray}
 V_L(g)&=&\int_{C_{\rm left}}\frac{dz}{2\pi i} g(z) v(z),\\
\label{eq:vz}
 v(z)&=& [\hQ,\,\tilde{v}(z)] = 
{1\over
\sqrt{2}}(c(z)\otimes
\sigma_3) J^1(z)+{1\over
\sqrt{2}}
\eta e^{\phi}\psi^1(z).
\end{eqnarray}
For the operators $V_L$ and $\tilde{V}_L$, we find the commutation
relations 
\begin{eqnarray}
&& [\hQ,\,\tilde{V}_L(g)]= V_L(g),\\
&& [\tilde{V}_L(g_1),\,V_L(g_2)]=-\frac{1}{2}C_L(g_1g_2)\otimes \sigma_3,
\end{eqnarray}
where $C_L(g)\equiv \int_{\rm C_{\rm left}}{dz\over 2\pi i}g(z)c(z)$.
Then, taking into account of the properties of these
operators associated with the star product \cite{KT2}, we can obtain
\begin{eqnarray}
&&e^{-\hat{\Phi}_0}*\hat{Q}_{\rm B} e^{\hat{\Phi}_0}
=(e^{\tilde{V}_L(F)}\hat{Q}_{\rm B} e^{-\tilde{V}_L(F)})I
=-V_L(F)I+{1\over 4}C_L(F^2)I\otimes \sigma_3.
\end{eqnarray}
The $\xi$ zero mode is not contained in both operators $V_L(F)$ and
$C_L(F^2)$ and the identity string field satisfies $\eta_0 I=0$.
As a result, we find that $\heta_0(e^{-\hat{\Phi}_0}*\hQ 
e^{\hat{\Phi}_0})=0$ and the equation of motion holds.

Concerning the vacuum energy, we can obtain it by calculating the
correlation function
$\Bra(\hat{\eta}_0\hat{\Phi}_0)(e^{-t\hat{\Phi}_0}\hat{Q}_{\rm B}
e^{t\hat{\Phi}_0})\Ket$.  
Similarly, it can be seen that there is no $\xi$ zero mode in 
$e^{-t\hat{\Phi}_0}\hat{Q}_{\rm B} e^{t\hat{\Phi}_0}$ for arbitrary $t$.
This fact is sufficient to
show that the vacuum energy of the classical solution
vanishes exactly in the same way as that
of the GSO$(+)$ solution in ref.~\cite{KT2},

\section{Superstring field theory around the solution}

We consider the action expanded around the classical solution in order
to provide physical interpretation of the solution. If we expand the
string field as $e^{\hat{\Phi}}=e^{\hat{\Phi}_0}e^{\hat{\Phi}'}$, the
action (\ref{Eq:action}) becomes
\begin{eqnarray}
\label{eq:expaction}
 S[\hat{\Phi};\,\hQ]=S[\hat{\Phi}_0;\,\hQ]+S[\hat{\Phi}';\,\hQ'].
\end{eqnarray}
The first term of the right-hand side corresponds to the vacuum energy
of the solution, which is seen to be zero as discussed above.
Then, the expanded action
takes the same form as the original action except that the BRS charge is
changed depending on the classical solution.
Accordingly, we will
investigate the new BRS charge $\hQ'$ to determine the spectrum around
the solution.

\subsection{Fermionization and rebosonization}

To find the spectrum around the classical solution, it is convenient to
fermionize the scalar field $X^9(z)$ as in
refs.~\cite{MajSen,Sen:1998ex}:
\begin{eqnarray}
\label{eq:rule1}
 e^{\pm {i\over \sqrt{2\alpha'}}X^9(z)}= {1\over
       \sqrt{2}}(\xi^9(z)\pm
 i\eta^9(z))\otimes \tau_1,
\end{eqnarray}
where $\xi^9(z)$ and $\eta^9(z)$ are fermionic fields and the Pauli
matrices $\tau_i$ denote cocycle factors. To ensure correct
(anti-)commutation relations between various fields, we also attach a
cocycle factor $\tau_3$ to all other fermionic fields. For example,
$\psi^9(z)$ is replaced with $\psi^9(z)\otimes \tau_3$, and the
derivation operator $\hat{\eta}_0$ is written as
$\hat{\eta}_0=\eta_0\otimes 
\sigma_3\otimes \tau_3$. Using the
fermionization 
rule (\ref{eq:rule1}), the supercurrents (\ref{eq:J1v2}),
(\ref{eq:J2v2}) and (\ref{eq:J3v2}) can be expressed as
\begin{eqnarray}
\label{eq:J1f}
 {\bf J}^1(z,\theta)&=&\eta^9(z)\otimes \sigma_2\otimes \tau_1+\theta\,
(-i\psi^9\xi^9(z))\otimes \sigma_1\otimes \tau_2\,,
\\
\label{eq:J2f}
 {\bf J}^2(z,\theta)&=&\xi^9(z)\otimes \sigma_2\otimes\tau_1
+\theta\,(i\psi^9\eta^9(z))\otimes \sigma_1\otimes \tau_2\,,\\
\label{eq:J3f}
 {\bf J}^3(z,\theta)&=&\psi^9(z)\otimes \sigma_3\otimes \tau_3
+\theta\,(-i\xi^9\eta^9(z))\otimes {\bf 1}\otimes {\bf 1}\,.
\end{eqnarray}
Similarly, the energy-momentum tensor (\ref{Eq:emtensor}) and the
world-sheet supercurrent (\ref{Eq:scurrent}) are rewritten as
\begin{eqnarray}
 T^9(z)&=&\left(-{1\over
2}\xi^9\partial\xi^9(z)-{1\over 2}\eta^9\partial\eta^9(z)-{1\over
2}\psi^9\partial\psi^9(z)\right)\otimes{\bf 1}\otimes{\bf 1}\,,\\
G^9(z)&=&-i\xi^9\eta^9\psi^9(z)\otimes \sigma_3\otimes \tau_3\,.
\end{eqnarray}
Then, the BRS charge is expressed as $\hQ=Q_{\rm B}\otimes
\sigma_3\otimes \tau_3$.

Since we compactify the 9-th direction to the circle, the momentum along
this direction is quantized and it is labeled by even and odd integers. 
Applying the fermionization rule to the string field, 
the GSO$(+)$ states with the odd momentum carry the cocycle factor
$\tau_1$. Since the GSO$(-)$ states correspond to fermionic vertex
operators, the cocycle factor $\tau_3$ ($\tau_2$) is attached to the
GSO$(-)$ states with even (odd) momentum. Then, we can express the
string field as
\begin{eqnarray}
\label{eq:hPhi}
 \hat{\Phi}=\Phi_+^{\rm e}\otimes {\bf 1}\otimes{\bf 1}
+\Phi_+^{\rm o}\otimes {\bf 1}\otimes \tau_1
+\Phi_-^{\rm e}\otimes  \sigma_1\otimes \tau_3
+\Phi_-^{\rm o}\otimes \sigma_1\otimes \tau_2,
\end{eqnarray}
where the subscript $\pm$ denotes GSO parity and
the superscript e (o) implies the state with even (odd)
momentum. For details, the world-sheet fermion number is defined as
\begin{eqnarray}
(-1)^{\hat{F}}|\Phi_{\pm }\rangle=\pm |\Phi_{\pm }\rangle\,,
\end{eqnarray}
where the operator $\hat{F}$ in our convention is\footnote{
Here $\phi$ is a bosonized ghost coming from $\gamma=\eta e^\phi$,\ 
$\beta=e^{-\phi}\partial \xi$, and $\psi^\mu\ (\mu=0,1,\cdots,9)$ are
matter fermions. The reader should
 not confuse them with $\phi^9$ in eq.~(\ref{eq:rule2})
and the lowest components $\psi^a$ of the $su(2)$ supercurrent 
${\bf J}^a(z,\theta)\ (a=1,2,3)$.}
\begin{eqnarray}
\label{eq:F}
&&
\hat{F}=\oint{dz\over 2\pi i}\left(\sum_{k=1}^5:\psi_+^k\psi_-^k:(z)
-\partial \phi(z)\right),\\
&&
\psi_{\pm}^1\equiv \frac{i}{\sqrt{2}}(\psi^0\pm \psi^1),\ \ 
\psi_{\pm}^k\equiv \frac{1}{\sqrt{2}}(\psi^{2k-2}\pm i \psi^{2k-1}),~~
k=2,3,4,5.
\end{eqnarray}
The momentum parity is defined as
\begin{eqnarray}
\label{eq:n}
&&(-1)^{\hat{n}}|\Phi^{\rm e}\rangle=+ |\Phi^{\rm
 e}\rangle\,,
~~~~(-1)^{\hat{n}}|\Phi^{\rm o}\rangle=- |\Phi^{\rm
 o}\rangle\,,
\end{eqnarray}
where the operator $\hat{n}$ counting the 9-th momentum is given by
\begin{eqnarray}
\label{eq:n1}
&&\hat{n}=\oint{dz\over 2\pi i}{i\over \sqrt{2\alpha'}}\partial X^9(z)
=\oint{dz\over 2\pi i}i\eta^9\xi^9(z)\,.
\end{eqnarray}
In general, the string field of an even ghost number is expanded by the
same cocycle factors. The string field of an odd ghost number, like
gauge transformation parameters, can be written as
\begin{eqnarray}
\label{eq:hlambda}
 \hat{\Lambda}&=&
\Phi_+^{{\rm e}}\otimes \sigma_3\otimes \tau_3
+\Phi_+^{{\rm o}}\otimes \sigma_3\otimes \tau_2
+\Phi_-^{{\rm e}}\otimes \sigma_2\otimes {\bf 1}
+\Phi_-^{{\rm o}}\otimes \sigma_2\otimes \tau_1.
\end{eqnarray}

We can find another representation of the conformal field theory for
$(\psi^9,\xi^9, \eta^9)$ by the rebosonization \cite{MajSen,Sen:1998ex}
\begin{eqnarray}
\label{eq:rule2}
 (\xi^9(z)\pm i\psi^9(z))= \sqrt{2} e^{\pm {i\over
  \sqrt{2\alpha'}}\phi^9(z)}\otimes \tilde{\tau}_1,
\end{eqnarray}
where the Pauli matrices $\tilde{\tau_i}$ are cocycle factors and we
assign the cocycle 
$\tilde{\tau}_3$ to fermionic fields except $\psi^9(z)$ and $\xi^9(z)$.
We can easily rewrite all operators and the string field using the
bosonization rule (\ref{eq:rule2}). In particular, the supercurrents 
~(\ref{eq:J1f}), (\ref{eq:J2f}) and (\ref{eq:J3f}) are expressed as
\begin{eqnarray}
\label{eq:J1p}
 {\bf J}^1(z,\theta)\!&\!=\!&\!
\eta^9\otimes \sigma_2\otimes \tau_1\otimes \tilde{\tau}_3
-\theta{i\over \sqrt{2\alpha'}}\partial \phi^9(z)
\otimes \sigma_1\otimes \tau_2\otimes{\bf 1},
\\
\label{eq:J2p}
 {\bf J}^2(z,\theta)\!&\!=\!&\!\sqrt{2}\cos\!\left(\!\phi^9\over
				      \sqrt{2\alpha'}\!\right)\!(z)
 \otimes \sigma_2\otimes \tau_1\otimes \tilde{\tau}_1
+\theta
\sqrt{2}\eta^9\sin\!\left(\!\phi^9\over
			\sqrt{2\alpha'}\!\right)\!(z)\otimes
\sigma_1\otimes \tau_2\otimes \tilde{\tau}_2,~\\
\label{eq:J3p}
 {\bf J}^3(z,\theta)\!&\!=\!&\!
\sqrt{2}\sin\!\left(\!\phi^9\over \sqrt{2\alpha'}\!\right)\!(z)
 \otimes \sigma_3\otimes \tau_3\otimes \tilde{\tau}_1
-\theta\sqrt{2}\eta^9\cos\!\left(\!\phi^9\over
			\sqrt{2\alpha'}\!\right)\!(z)\otimes
{\bf 1}\otimes {\bf 1}\otimes \tilde{\tau}_2.
\end{eqnarray}

We note that we have to change the normalization of the action if we
apply the fermionization or 
the rebosonization to the string field. In the action 
(\ref{Eq:action}), we take the trace of all Chan-Paton indeces. If we
fermionize $X^9$, the Chan-Paton factors $\tau_i$ with their trace
produce an extra factor of two for the action. Consequently, we must
divide the action by two in order to provide the same action for the
component fields. Furthermore, if we rebosonize and introduce the
additional Chan-Paton factors $\tilde{\tau}_i$, we need to divide the
action by four.

\subsection{The theory expanded around the solution}

The new BRS operator in the expanded action $S[\hat{\Phi}';\hat{Q}'_{\rm
B}]$ in (\ref{eq:expaction}) around 
a solution $\hat{\Phi}_0$ to the equation of motion (\ref{Eq:eom})
 is generically expressed as
\begin{eqnarray}
\label{eq:hQpgen}
 \hQ'\hat{\Psi}=
\hQ\hat{\Psi}+\hat{A}_0*\hat{\Psi}
-(-)^{{\rm gh}({\hat{\Psi}})}\hat{\Psi}*\hat{A}_0, 
\ \ \ \hat{A}_0=e^{-\hat{\Phi}_0}*\hQ e^{\hat{\Phi}_0}
\ \ \ \ {\rm for}\ {}^\forall \hat{\Psi}.
\end{eqnarray}
This formula can be derived as appendix B in ref.~\cite{KT2},
for example,
because algebraic relations are almost the 
 same as the original GSO projected theory \cite{BSZ}.
Here ${\rm gh}(\hat{\Psi})$ denotes ghost number
of $\hat{\Psi}$ and it is counted by $n_{\rm gh}=-\oint \frac{dz}{2\pi
i}(:\! bc\! :+:\!\xi\eta\!:)$.
A string field $\hat{\Psi}$ takes the form of
 (\ref{eq:hPhi}) for even ghost number
and (\ref{eq:hlambda}) for odd ghost number.
The sign factor $(-)^{{\rm gh}(\hat{\Psi})}$ instead of ``Grassmannality''
appears because $\hat{Q}'_{\rm B}$ should be an anti-derivation as
original $\hat{Q}_{\rm B}$
in the sense that $\hat{Q}'_{\rm B}(\hat{\Psi}_1*\hat{\Psi}_2)
=(\hat{Q}'_{\rm B}\hat{\Psi}_1)*\hat{\Psi}_2+
(-)^{{\rm gh}(\hat{\Psi}_1)}\hat{\Psi}_1*(\hat{Q}'_{\rm B}\hat{\Psi}_2)$
\cite{Ohmori:2002kj}.

We rewrite the operator $\tilde{v}(z)$ in  
the solution (\ref{eq:solution_gen1}) 
by using the fermionic fields $(\psi^9,\xi^9,\eta^9)$ through the 
fermionization rule (\ref{eq:rule1}):
\begin{eqnarray}
\label{eq:vtila_gm1}
 \tilde{v}(z)
&=&{1\over \sqrt{2}}c\xi
 e^{-\phi}\,\eta^9(z)
\otimes\sigma_1\otimes\tau_2.
\end{eqnarray}
The operator (\ref{eq:vz}) can be written as
\begin{eqnarray}
v(z)=
\left({-i\over \sqrt{2}}c\psi^9\xi^9(z)+
{1\over
\sqrt{2}}
\eta e^{\phi}\eta^9(z)\right)\otimes
\sigma_2\otimes\tau_1. 
\end{eqnarray}
The operator $\hQ'$ (\ref{eq:hQpgen}) for the solution (\ref{eq:vtila_gm1})
can be found as
\begin{eqnarray}
 \hQ'=(\Q+\frac{1}{4}C(F^2))\otimes \sigma_3\otimes \tau_3
-V_L(F)-(-1)^{\hat{F}+\hat{n}}V_R(F),
\end{eqnarray}
where 
$C(F^2)=C_L(F^2)+C_R(F^2)$,
$C_R(g)\equiv\int_{C_{\rm right}}{dz\over 2\pi i}g(z)c(z)$,
$V_R(g)\equiv\int_{C_{\rm right}}{dz\over 2\pi i}g(z)v(z)$, 
$C_{\rm
right}$ is a counter-clockwise path along 
a half of the unit circle:($|z|=1,\,{\rm Re}\,z<0$)
as in ref.~\cite{KT2}, 
and the operators $\hat{F}$ and $\hat{n}$ are given by
eqs.~(\ref{eq:F}) and (\ref{eq:n1}).
The extra sign factor $(-1)^{\hat{F}+\hat{n}}$ in front of $V_R(F)$
comes from exchange of order of $\hat{\Psi}$ and $v(z)$
in eq.~(\ref{eq:hQpgen}).
This new BRS operator can be rewritten in terms of a similarity
transformation from the original operator,
\begin{eqnarray}
\label{eq:newBRST}
\hQ'
&=&e^{\tilde{V}_L(F)+(-1)^{\hat{F}+\hat{n}}\tilde{V}_R(F)
}\,\hQ\,e^{-\tilde{V}_L(F)-(-1)^{\hat{F}+\hat{n}}
\tilde{V}_R(F)}.
\end{eqnarray}
We notice that this relation cannot be used for a field redefinition 
in the expanded action because of
$[\hat{\eta}_0,\tilde{V}_L(F)+(-1)^{\hat{F}+\hat{n}}\tilde{V}_R(F)]\ne 0$.

Furthermore, we can find another expression of the new BRS charge in
terms of $(\phi^9,\eta^9)$. If
the new BRS charge acts on the state of $(-1)^{\hat{F}+\hat{n}}=+1$, it
becomes 
\begin{eqnarray}
\label{eq:newBRSTphi1}
\hQ'&=&
e^{-{i\over 2\sqrt{\alpha'}}(\phi^9_L(F)+
\phi^9_R(F))\otimes \sigma_1
\otimes\tau_2}\,\hQ\,
e^{{i\over 2\sqrt{\alpha'}}(\phi^9_L(F)+
\phi^9_R(F))\otimes \sigma_1
\otimes\tau_2},
\end{eqnarray}
and for the case of $(-1)^{\hat{F}+\hat{n}}=-1$,
\begin{eqnarray}
\label{eq:newBRSTphi2}
\hQ'&=&
e^{-{i\over 2\sqrt{\alpha'}}(\phi^9_L(F)-
\phi^9_R(F))\otimes \sigma_1
\otimes\tau_2}\,\hQ\,
e^{{i\over 2\sqrt{\alpha'}}(\phi^9_L(F)-
\phi^9_R(F))\otimes \sigma_1
\otimes\tau_2},
\end{eqnarray}
where $\phi^9_{L/R}(F)\equiv\int_{C_{\rm left/right}}\frac{dz}{2\pi i}
F(z) \phi^9(z)$.
They are derived from the direct calculation or from the expression
(\ref{eq:newBRST}) and the following anti-commutation relation,
\begin{eqnarray}
 \{\hQ,\,\Omega_{L/R}(F)\}&=&
   2\sqrt{\alpha'}\tilde{V}^1_{L/R}(F)+i\phi^9_{L/R}(F)\otimes \sigma_1
\otimes\tau_2\,,\\
\Omega_{L/R}(F) &\equiv& 
-\int_{C_{\rm left/right}}\frac{dz}{2\pi i}
F(z) \,i\,c\xi \partial \xi e^{-2\phi}\,\phi^9(z)
\otimes \sigma_2\otimes\tau_1.
\end{eqnarray}

Noting $[\hat{\eta}_0,\phi^9_{L/R}(F)\otimes \sigma_1\otimes\tau_2]=0$,
these expressions given in eqs.~(\ref{eq:newBRSTphi1}) and
(\ref{eq:newBRSTphi2}) for the new BRS operator
 imply that the expanded
action around the solution can be transformed back to the original
action by the string field redefinition,
\begin{eqnarray}
\label{eq:sineredef}
 \hat{\Phi}''&=&e^{{i\over 2\sqrt{\alpha'}}
\phi^9_L(F)I\otimes \sigma_1
\otimes\tau_2} *\hat{\Phi}' *e^{-{i\over 2\sqrt{\alpha'}}
\phi^9_L(F)I\otimes \sigma_1
\otimes\tau_2}.
\end{eqnarray}
Actually, this string field redefinition does not change the interaction
terms in the action and,
depending on the $(-1)^{\hat{F}+\hat{n}}$ parity of the string
field, the redefinition can be rewritten as
\begin{eqnarray}
\label{eq:sineredef2}
 \hat{\Phi}''
&=&\left\{
\begin{array}[tb]{cc}
e^{{i\over 2\sqrt{\alpha'}}(\phi^9_L(F)+
\phi^9_R(F))\otimes \sigma_1
\otimes\tau_2}\hat{\Phi}' &
\mbox{on}~~(-1)^{\hat{F}+\hat{n}}=+1 \\
e^{{i\over 2\sqrt{\alpha'}}(\phi^9_L(F)-
\phi^9_R(F))\otimes \sigma_1
\otimes\tau_2}\hat{\Phi}' &
\mbox{on}~~(-1)^{\hat{F}+\hat{n}}=-1.
\end{array}
\right.
\end{eqnarray}
The difference in sign in the right-hand side arises from
(anti-)commutation 
relations of Chan-Paton factors. For the string field (\ref{eq:hPhi}),
the $(-1)^{\hat{F}+\hat{n}}=+1$ sector involves cocycle factors ${\bf
1}\otimes {\bf 
1}$ and $\sigma_1\otimes \tau_2$, which commute with
the generator $\sigma_1\otimes \tau_2$ 
of the string field redefinition. However, the generator anti-commutes
with the cocycle factors ${\bf 1}\otimes \tau_1$ and $\sigma_1\otimes
\tau_3$ and then the minus sign appears for the
$(-1)^{\hat{F}+\hat{n}}=-1$ sector, $\Phi_+^{\rm o}$ and $\Phi_-^{\rm
e}$ in eq.~(\ref{eq:hPhi}).

Though the expanded action is transformed to the original one, 
the string field redefinition has a physical effect.
As discussed for the case of the Wilson line solution in
ref.~\cite{KT2}, the 
spectrum is changed from that of the original theory due to 
the zero-mode of the operator $\phi^9(z)$. 
We have no zero-mode in the
operator $\phi^9(F)\equiv\phi^9_L(F)+\phi^9_R(F)$ because
we impose the condition $F(-1/z)=z^2F(z)$ 
in the solution (\ref{eq:solution_gen1}) and then the coefficients of
the zero-mode cancel as
\begin{eqnarray}
\int_{C_{\rm left}}{dz\over 2\pi i}F(z)+
\int_{C_{\rm right}}{dz\over 2\pi i}F(z)=0\,,
\end{eqnarray}
whereas $\phi_\Delta^9(F)\equiv \phi^9_L(F)-\phi^9_R(F)$ includes the
zero-mode. 
As a result, the $(-1)^{\hat{F}+\hat{n}}=-1$ sector is multiplied by the
extra factor,
\begin{eqnarray}
\label{eq:factor_f}
 \exp\left(i\frac{f}{\sqrt{\alpha'}}\,
\hat{\phi}^9_0\otimes \sigma_1\otimes \sigma_2\right),\ \ \ 
f\equiv\int_{C_{\rm left}}{dz\over 2\pi i}F(z),
\end{eqnarray}
where $\hat{\phi}^9_0$ denotes the zero-mode operator of $\phi^9(z)$.
This zero-mode factor changes the momentum of the string field along the
$\phi^9$ direction 
as $p_{\phi^9}\rightarrow p_{\phi^9}+f/\sqrt{\alpha'}$. The momentum
shift only for the $(-1)^{\hat{F}+\hat{n}}=-1$ sector is exactly the
same effect as that of a tachyonic lump solution as discussed in
the context of boundary conformal field theory \cite{Sen:1998ex}. Hence,
our analytic solution (\ref{eq:solution_gen1}) represents the same
tachyonic lump solution in open superstring field theory, and the half
integration mode, $f$, corresponds to the Wilson line
along the $\phi^9$ direction. 

\subsection{The expanded theory at the critical value of $f$}

We discuss a tachyonic lump solution corresponding to the critical
value of $f$ in (\ref{eq:factor_f}), namely
\begin{eqnarray}
 f=\frac{2m+1}{\sqrt{2}},\ \ \ m\in {\bf Z}.
\end{eqnarray}
At the critical value, the redefined field $\hat{\Phi}''$ 
(\ref{eq:sineredef2}) can be rewritten 
again by the fermionic fields $(\psi^9,\xi^9,\eta^9)$ instead of
$(\phi^9,\eta^9)$. Moreover, we can write its string field by the
original string coordinates $(X^9,\psi^9)$ through the rebosonization of
$(\xi^9,\eta^9)$ to $X^9$.

Using the fermionic fields $(\psi^9,\xi^9,\eta^9)$, we can write the
redefined field as
\begin{eqnarray}
 \hat{\Phi}''
&=&e^{{i\over 2\sqrt{\alpha'}}\phi^9(F)\otimes \sigma_1
\otimes\tau_2}(\Phi_+^{{\rm e}}\otimes {\bf
 1}\otimes {\bf 1}
+\Phi_-^{{\rm o}}\otimes \sigma_1\otimes \tau_2)\nonumber\\
&&+e^{{i\over 2\sqrt{\alpha'}}\phi^9_{\Delta}(F)\otimes \sigma_1
\otimes\tau_2}(\Phi_+^{{\rm o}}\otimes {\bf 1}\otimes \tau_1
+\Phi_-^{{\rm e}}\otimes \sigma_1\otimes \tau_3)\nonumber\\
&=&
\left(\cos\left(\frac{\phi^9(F)}{2\sqrt{\alpha'}}\right)
\Phi_+^{\rm e}
+i\sin\left(\frac{\phi^9(F)}{2\sqrt{\alpha'}}\right)
\Phi_-^{\rm o}\right)\otimes {\bf 1}\otimes {\bf 1}\nn
&&
+\left(i\sin\left(\frac{\phi^9(F)}{2\sqrt{\alpha'}}\right)
\Phi_+^{\rm e}
+\cos\left(\frac{\phi^9(F)}{2\sqrt{\alpha'}}\right)
\Phi_-^{\rm o}\right)\otimes \sigma_1\otimes \tau_2\nn
&&
+\left(\cos\left(\frac{\phi_\Delta^9(F)}{2\sqrt{\alpha'}}\right)
\Phi_+^{\rm o}
-\sin\left(\frac{\phi_\Delta^9(F)}{2\sqrt{\alpha'}}\right)
\Phi_-^{\rm e}\right)\otimes {\bf 1}\otimes \tau_1\nn
&&
\label{eq:redefSF3}
+\left(\sin\left(\frac{\phi_\Delta^9(F)}{2\sqrt{\alpha'}}\right)
\Phi_+^{\rm o}
+\cos\left(\frac{\phi_\Delta^9(F)}{2\sqrt{\alpha'}}\right)
\Phi_-^{\rm e}\right)\otimes \sigma_1\otimes \tau_3.
\end{eqnarray}
Here, the operators, $\cos(\phi^9_{(\Delta)}(F)/2\sqrt{\alpha'})$ and
$\sin(\phi^9_{(\Delta)}(F)/2\sqrt{\alpha'})$,
are represented in terms of the fermionic fields $\xi^9$ and $\psi^9$
instead of $\phi^9$ thanks to  (\ref{eq:rule2}) and
$\partial
\phi^9=-\sqrt{2\alpha'}\xi^9\psi^9$.

At first, we consider the operators
$\cos(\phi^9_\Delta(F)/2\sqrt{\alpha'})$  and
$\sin(\phi^9_\Delta(F)/2\sqrt{\alpha'})$.
When we introduce the operator counting $\phi^9$ momenta as
$\displaystyle \hat{n}_{\phi^9}=\oint \frac{dz}{2\pi i}i\partial
\phi^9/\sqrt{2\alpha'}$, the operators appeared in the redefinition
have $(-1)^{\hat{n}_{\phi^9}}=-1$  because they carry
the $\phi^9$ momentum
$p_{\phi^9}=f/\sqrt{\alpha'}=(2m+1)/\sqrt{2\alpha'}$. 
In addition, as in refs.~\cite{Sen:1998ex,MajSen}, we define the
operator $\hat{F}_{\phi^9}$  counting the fermion number of $\eta^9$ and
other spectator fermions, $\psi^\mu\ (\mu=0,1,\cdots,8)$ and
$e^{q\phi}$ ($q$:odd).
With this definition, the operators  
have $(-1)^{\hat{F}_{\phi^9}}=+1$. Then, we find that
the operators have $(-1)^{\hat{F}_{\phi^9}+\hat{n}_{\phi^9}}=-1$.

Next, we consider the original fermion number and momenta of the
operators. If we change the sign of $\phi^9$,
the fermions $\xi^9$, $\psi^9$ are transformed as
\begin{eqnarray}
 \xi^9 \,\rightarrow\, \xi^9,
\ \   \psi^9 \,\rightarrow\, -\psi^9,
\end{eqnarray}
because $\phi^9$ is related to $\xi^9$ and $\psi^9$ through the
rebosonization rule (\ref{eq:rule2}). Therefore we can determine the
$(-1)^{\hat{F}}$ parity of some operators by means of the parity
transformation of $\phi^9$. We can find that
$\cos(\phi^9_\Delta(F)/2\sqrt{\alpha'})$
has $(-1)^{\hat{F}}=+1$ and $\sin(\phi^9_\Delta(F)/2\sqrt{\alpha'})$
has $(-1)^{\hat{F}}=-1$.
 As discussed in ref.~\cite{Sen:1998ex}, we have the relation:
\begin{eqnarray}
 (-1)^{\hat{F}}(-1)^{\hat{n}}=
 (-1)^{\hat{F}_{\phi^9}}(-1)^{\hat{n}_{\phi^9}}.
\end{eqnarray}
Combining these results, we can determine the values of
$(-1)^{\hat{F}}$ and  $(-1)^{\hat{n}}$ individually for these
operators. 
The resulting parities for operators are listed in the following table:
\begin{eqnarray}
 \begin{array}{c|c|c|c|c}
  & (-1)^{\hat{F}}& (-1)^{\hat{n}}
  & (-1)^{\hat{F}_{\phi^9}}& (-1)^{\hat{n}_{\phi^9}}\\
\hline
\hline
{\displaystyle
 \cos\left(\frac{\phi_\Delta^9(F)}{2\sqrt{\alpha'}}\right)} &
{\rm +} & {\rm -} & {\rm +} & {\rm -} \\
\hline
{\displaystyle
 \sin\left(\frac{\phi_\Delta^9(F)}{2\sqrt{\alpha'}}\right)} &
{\rm -} & {\rm +} & {\rm +} & {\rm -} \\
 \end{array}
\end{eqnarray}
Based on a similar consideration, we have the following results for
other operators:
\begin{eqnarray}
 \begin{array}{c|c|c|c|c}
  & (-1)^{\hat{F}}& (-1)^{\hat{n}}
  & (-1)^{\hat{F}_{\phi^9}}& (-1)^{\hat{n}_{\phi^9}}\\
\hline
\hline
{\displaystyle
 \cos\left(\frac{\phi^9(F)}{2\sqrt{\alpha'}}\right)} &
{\rm +} & {\rm +} & {\rm +} & {\rm +} \\
\hline
{\displaystyle
 \sin\left(\frac{\phi^9(F)}{2\sqrt{\alpha'}}\right)} &
{\rm -} & {\rm -} & {\rm +} & {\rm +} \\
 \end{array}
\end{eqnarray}

From these results, we find that the first and second terms in
eq.~(\ref{eq:redefSF3}) have $(-1)^{\hat{F}}=+1$ and $(-1)^{\hat{n}}=+1$
 and the third and fourth have  $(-1)^{\hat{F}}=-1$ and
 $(-1)^{\hat{n}}=-1$, and then all components in the redefined
string field (\ref{eq:redefSF3}) have $(-1)^{\hat{F}+\hat{n}}=+1$. Before 
the redefinition, the fields with ${\bf 1}\otimes \tau_1$ and $\sigma_1
\otimes \tau_3$ have $(-1)^{\hat{F}+\hat{n}}=-1$.
The parity of these states is
changed after the redefinition. Alternatively, the quantum number of
$\hat{F}+\hat{n}$ can be regarded as the 
fermion number assigning $+1$ to the fields $\psi^9$, $\xi^9$, $\eta^9$
in the fermionic representation. Consequently, with the fermionic 
representation, the statistical property of
the fields with ${\bf 1}\otimes \tau_1$ and $\sigma_1
\otimes \tau_3$ are changed under the string field redefinition.
In order to ensure correct (anti-)commutation relations, we have to
assign a cocycle factor of $\tilde{\tau}_1$ to these fields, and assign
a cocycle factor $\tilde{\tau}_3$ to the derivations $\hQ$ and $\heta_0$.
After all, the redefined string field can be expressed using the fermionic
representation as
\begin{eqnarray}
\label{eq:redefSF}
 \hat{\Phi}''=\Psi_+^{\rm e} \otimes{\bf 1}\otimes {\bf 1}\otimes {\bf
  1} 
+\Psi_+^{\prime {\rm e}}\otimes {\bf 1}\otimes \tau_1\otimes \tilde{\tau}_1
+\Psi_-^{\rm o}\otimes \sigma_1\otimes \tau_3\otimes \tilde{\tau}_1
+\Psi_-^{\prime {\rm o}}\otimes \sigma_1\otimes \tau_2\otimes {\bf 1}.
\end{eqnarray}

Now, let us express the string field (\ref{eq:redefSF}) in terms of
the fields $(X^9,\psi^9)$ through the rule (\ref{eq:rule1}).
When we rebosonize $(\xi^9,\eta^9)$ to $X^9$, we have to assign a
cocycle factor $\tau_1$ to states with an odd momentum and retain the
cocycle factors $\tau_i$ under the earlier fermionization.  According to
this procedure,
the string fields $\Psi_-^{\rm o}$ and $\Psi_-^{\prime {\rm o}}$ acquire an
additional cocycle factor of $\tau_1$ under the rebosonization. Hence,
with the fields $(X^9,\psi^9)$, the string field can be rewritten as
\begin{eqnarray}
\label{eq:redefSF2}
 \hat{\Phi}''=\Psi_+^{\rm e} \otimes{\bf 1}\otimes {\bf 1}\otimes {\bf
  1} 
+\Psi_+^{\prime {\rm e}}\otimes {\bf 1}\otimes \tau_1\otimes \tilde{\tau}_1
+\Psi_-^{\rm o}\otimes \sigma_1\otimes \tau_2\otimes \tilde{\tau}_1
+\Psi_-^{\prime{\rm o}}\otimes \sigma_1\otimes \tau_3\otimes {\bf 1}.
\end{eqnarray}
The derivations are expressed as same as before:
\begin{eqnarray}
\label{eq:derivations}
 \hQ=\Q\otimes \sigma_3\otimes \tau_3\otimes \tilde{\tau}_3,
\ \ \ 
 \hat{\eta}_0=\eta_0\otimes \sigma_3\otimes \tau_3\otimes \tilde{\tau}_3.
\end{eqnarray}
Here, we write the cocycle factors appeared in the string field
(\ref{eq:redefSF2}) and the derivations (\ref{eq:derivations}) as
\begin{eqnarray}
 \Sigma_3={\bf 1}\otimes \tau_1\otimes \tilde{\tau}_1,\ \ 
 \Sigma_1=\sigma_1\otimes \tau_2\otimes \tilde{\tau}_1,\ \ 
 \Sigma_2=\sigma_1\otimes \tau_3\otimes {\bf 1},\ \ 
 \sigma=\sigma_3\otimes \tau_3\otimes \tilde{\tau}_3.
\end{eqnarray}
These matrices satisfy the following relations:
\begin{eqnarray}
&&\Sigma_1^2=\Sigma_2^2=\Sigma_3^2=\sigma^2=1\,,\\
&&[\sigma,\Sigma_3]=\{\sigma,\Sigma_1\}=
\{\sigma,\Sigma_2\}=0\,,\\
&&\Sigma_i\Sigma_j=i\epsilon_{ijk}\Sigma_k\,,~~~(i\ne j)\,. 
\end{eqnarray}
We can represent the same algebra by the alternative Pauli matrices
$\sigma_i$ and $\tau_i$: 
\begin{eqnarray}
 \Sigma'_1= \sigma_1\otimes \tau_1,\ \ 
 \Sigma'_2= \sigma_1\otimes \tau_2,\ \ 
 \Sigma'_3= {\bf 1}\otimes \tau_3,\ \ 
 \sigma'= \sigma_3\otimes {\bf 1}.
\end{eqnarray}
Therefore, we can identify $(\Sigma_i', \sigma')$ with $(\Sigma_i,\sigma)$
if we divide the action by two to compensate their different 
normalization. 

Finally, under the above identification, we can represent the redefined
string field in terms of $(X^9,\psi^9)$ as
\begin{eqnarray}
\label{eq:redefSF3_2}
 \hat{\Phi}''=\Psi_+^{\rm e} \otimes {\bf 1}\otimes{\bf 1}
+\Psi_+^{\prime{\rm e}}\otimes {\bf 1}\otimes \tau_3
+\Psi_-^{\rm o}\otimes \sigma_1\otimes \tau_1
+\Psi_-^{\prime{\rm o}}\otimes \sigma_1\otimes \tau_2,
\end{eqnarray}
and the derivations as
\begin{eqnarray}
\label{eq:derivations3}
 \hQ=\Q\otimes \sigma_3\otimes {\bf 1},
\ \ \ 
 \hat{\eta}_0=\eta_0\otimes \sigma_3\otimes {\bf 1}.
\end{eqnarray}
The resulting string field theory including this string field and the
derivations is exactly the same 
theory on a D-brane-anti-D-brane pair discussed in ref.~\cite{BSZ}, in
which $\sigma_i$ are the internal Chan-Paton indices to include the
GSO$(-)$ sector and $\tau_i$ 
correspond to the conventional Chan-Paton indices introduced for a pair
of branes. The string fields connecting a D-brane and an anti-D-brane
have odd momenta along the $X^9$ direction, and the string fields attached
both ends of the string to a single brane have even
momenta. Consequently,
the string field theory at the critical value of $f$
describes the D-brane-anti-D-brane system in which a D-brane and an
anti-D-brane are situated at antipodal points along the circle with the
critical radius in the T-dual picture.

Thus, we find that the tachyonic lump solution corresponding to the
critical value of $f$ changes the theory on a single non-BPS D-brane to that of
a D-brane-anti-D-brane pair. This is physically the same result obtained
before in terms of boundary conformal field theory
\cite{Sen:1998ex}. But, we should comment on a superficial
difference between these results. In our case, the 
resulting branes are put on the $X^9$ direction, while in
ref.~\cite{Sen:1998ex} the branes are on the
direction represented by $\phi^{\prime 9}(z)$, which is another bosonic field
given by a rebosonization of $(\eta^9,\psi^9)$.
As discussed in the previous section, the theory possesses the
$su(2)$ supercurrent algebra but the $SU(2)$ symmetry is broken on a
non-BPS D-brane. However, it turns out that the $SU(2)$ symmetry is
restored in the NS 
sector of the theory with the critical value of $f$ and the bosonic
coordinates $(X^9,\phi^9,\phi^{\prime 9})$ can be rotated under this
symmetry because all sectors have  $(-1)^{\hat{F}+\hat{n}}=+1$
after the string field redefinition around the solution. 
Therefore, the difference is resolved by the $SU(2)$ rotation
of $X^9$ to $\phi^{\prime 9}$. 

\section{Concluding remarks}

We constructed the analytic classical solution in superstring field
theory on the non-BPS D-brane in which the one direction $X^9$ is
compactified to a 
circle with the critical radius. The solution corresponds to the
tachyonic lump solution which corresponds to the Wilson line along the
$\phi^9$ direction. The vacuum energy of the solution vanishes exactly
as that of the BPS case.
At the critical value of $f$, the theory expanded around
the solution is equivalent to the theory on a
D-brane-anti-D-brane pair, including the interaction terms. These
results agree with the facts expected from boundary
conformal field theory. The $su(2)$ supercurrent algebra was useful for
the analyses of the solution.

In ref.~\cite{KT2}, we found some features of the solution on BPS
D-branes. The solution has a well-defined Fock space expression and 
the half integration mode $f$ is invariant under a class of 
gauge transformations in superstring field theory
but other modes are not.
Employing the same technique in ref.~\cite{KT2}, we can easily find that
the same is true in the case of non-BPS D-branes.

We should discuss the Ramond sector, which was out of the scope of this
paper, to complete the correspondence of our solution to the tachyonic
lump. The action on non-BPS D-branes including the Ramond sector is
supposed to be constructed by extending the action on BPS D-branes given
by ref.~\cite{rf:Michi}. In the extended theory including the Ramond
sector, our solution will satisfy the equation of motion. The problem
is whether the string field redefinition, especially at the critical
value of $f$, reproduces the expected result of the Wilson line along the
$\phi^9$ direction. 
It seems complicated to incorporate GSO$(-)$ states in the Ramond sector
and assign appropriate cocycle factors consistently.

We can apply our method constructing the analytical solution to other
cases of marginal deformations; a solution on non-BPS D-branes on an
orbifold \cite{Sen:1998ex} and a vortex solution on a
D-brane-anti-D-brane pair \cite{MajSen}. To realize marginal
deformations, we have to take the critical radius of the compactified
direction and the vacuum energy is always to be zero for these cases.
If we deform the radius away from the critical value, 
we may be able to find more
general solutions with a non-trivial vacuum energy. 
This problem is
interesting because such a general solution may teach us how the
closed string moduli changing the radius includes in open string field.

In this paper we show that there exists an analytic solution taking
the value in the GSO$(-)$ sector. This fact indicates the possible
existence of the analytic tachyon vacuum solution, at which non-BPS
D-branes completely disappear, in open superstring field theory. 
If we find the analytic solution, we could prove the non-existence of
open strings and the exact cancellation of the vacuum energy, as
discussed in the bosonic case \cite{rf:KT,Takahashi:2003xe,rf:tomo}.
We expect that the evaluation of the vacuum energy in the supersymmetric
case sheds lights on the problem what sort of regularization should be
applied to the bosonic theory.

\section*{Acknowledgements}

The authors would like to thank Yuji Igarashi and Katsumi Itoh for
useful discussions and the Yukawa Institute for Theoretical Physics at
Kyoto University for providing a stimulating atmosphere. Discussions
during the YITP workshop YITP-W-05-08 on ``String Theory and Quantum
Field Theory" were useful to complete this work. They are also grateful
to ``Summer Institute String Theory 2005" at Sapporo during which part of
this work was done.
I.~K. wishes to express his gratitude to Seiji Terashima for valuable
comments.


\end{document}